# A rule-based system proposal to aid in the evaluation and decision-making in external beam radiation treatment planning

R. C. Fernandes*, T. M. Machado*, H. J. Onisto*, A. D. Muñoz*, R. O. Silva*, L. R. Domingues*, G. C. Fonseca*, J. E. Bertuzzo*, M. T. Pereira†, B. Biazotto†, E. T. Costa†

**Department of Services & Hardware Engineering, Eldorado Research Institute – Campinas, SP, Brazil,*
*†Center for Biomedical Engineering, University of Campinas – Campinas, SP, Brazil*

## ABSTRACT

**As part of a plan launched by the Ministry of Health of Brazil to increase the availability of linear accelerators for radiotherapy treatment for the whole country, for which Varian Medical Systems company has won the bidding, a technical cooperation agreement was signed inviting Brazilian Scientific and Technological Institutions to participate in a technology transfer program. As a result, jointly, the Eldorado Research Institute and the Center for Biomedical Engineering of the University of Campinas presents in this work, the concepts behind of a proposed rule engine to aid in the evaluation and decision-making in radiotherapy treatment planning. Normally, the determination of the radiation dose for a given patient is a complex and intensive procedure, which requires a lot of domain knowledge and subjective experience from the oncologists' team. In order to help them in this complex task, and additionally, provide an auxiliary tool for less experienced oncologists, it is presented a project conception of a software system that will make use of a hybrid data-oriented approach. The proposed rule engine will apply both inference mechanism and expression evaluation to verify and accredit the quality of an external beam radiation treatment plan by considering, at first, the 3D-conformal radiotherapy (3DCRT) technique.**

## A. INTRODUCTION

Cancer, considered one of the most worldwide problem in oncology healthcare system, is a coined term used to refer to a large set of diseases that share as a main characteristic the disordered growth of cells, that usually invades tissues and organs, and which may eventually can spread out to other regions of the body. The source of its cause is diverse, and as such, may be related to habits and lifestyle, infectious agents, sociocultural and geographical environment, genetic predisposition, among others [1, 2, 3].

Within the Brazilian reality, according to the National Institute of Cancer of Brazil [In Portuguese: Instituto Nacional de Câncer José Alencar Gomes da Silva – INCA] the estimate for 2012/2013 was around 518 thousands new cases of cancer. Of which, near 258 thousand were men whilst 260 thousand were women, being their predominant occurrence in age groups over 50 years. The data also indicates that cancer represents the second cause of death in the Brazilian population. Notwithstanding, in another study, a projection by the International Agency for Research Cancer (IARC) indicates that, in Brazil, in 2030, there will be more than 733 thousands new cases of cancer, excepting those of non-melanoma skin. As a result, it represents an increase of approximately 54% compared to the same data from 2015 [4].

Radiation treatment planning is a very intensive interactive process composed of several activities in an image-guided scheme (e.g. CT, PET and/or MRI image slices) for devising an appropriate external beam radiotherapy or internal brachytherapy treatment for a patient with cancer. The objective is to provide a safe, accurate, easy-to-use tool for the clinical planning in order to maximize the radiation dose to the tumour (target volume) and minimize it to the surrounding healthy tissues [5].

Commonly, a team consisting of physician oncologists, medical physicists and dosimetrists carry out this process with the aid of software specifically tailored to serve as a clinical decision-support system for radiotherapy treatment. Those software makes possible for the oncologist team to executes sequences of target contours in the tissue region of interest, determine the beam field and calculates the radiation dose to be delivered to the patient, just to mention a few, for a chosen clinical modality (e.g. 3D-conformal radiotherapy – 3DCRT, Intensity-modulated radiation therapy – IMRT or Volumetric modulated arc radiotherapy – VMAT).

## B. The Technology Transfer Program Context

In order to expand the offer of radiotherapy treatment within the public healthcare system of Brazil (In Portuguese: Sistema Único de Saúde – SUS), the Brazilian Ministry of Health has launched a plan in 2012 that included the installation of up to 80 linear accelerators equipment throughout the country. Additionally, as part of the agreement between the Brazilian Ministry of Health and Varian Medical System, the Radiotherapy Public Call Notice nº 001/2016 invited the Brazilian Scientific and Technological Institutions [In Portuguese: Instituições Científicas e Tecnológicas – ICTs] interested to participate in a technical cooperation with Varian Medical Systems for a technology transfer program.

Stemming from this events, jointly, the Eldorado Research Institute [In Portuguese: Instituto de Pesquisas Eldorado (ELDORADO)] and the University of Campinas, through its Center for Biomedical Engineering [In Portuguese: Universidade Estadual de Campinas, através do Centro de Engenharia Biomédica – CEB-UNICAMP] formed an ICT, and as such, were selected to participate in two programs. The first, named, Training in Embedded Software Engineering for Linear Accelerators, held in Palo Alto, California, in the USA, and the second, called, 3D Treatment Planning, in Helsinki – Finland. Therefore, the present article is the result of the challenge launched in the aforementioned Radiotherapy Public Call for the second project to conceive ways to simplify/improve the 3D conformal radiotherapy treatment planning in the context of Brazil and, more specifically, for the SUS.

## C. The Proposed Project and its Goals

In oncology, the radiotherapy treatment planning is a very complex task due to the difficulties imposed for planning the radiation dose to delivery, or even to defining the contouring area and volume of an organ to receive the radiation without harming the surround ones for a given patient. As a result, the decision-making becomes a complex task because it requires a lot of domain knowledge and subjective experience of the oncologists and medical physicists in the front of clinical practice [6].

Although there are a plenty of clinical practice guidelines that are disease-specific recommendations to support clinical decision-making in accordance with the best evidence, usually, they are not often integrated into the front lines of care [7]. Additionally, with the amount of healthcare information available around of the world, produced by clinical trials (either recent papers discussing results, or ongoing clinical trials), and case series, beside of academy, industry and regulatory institutions, make the task of being up-to-date practically impossible for the professionals working on the front of clinical practice.

To deal with those challenges, looking at the Brazilian oncology ecosystem, the ELDORADO and CEB-UNICAMP proposes the project of developing a rule-based system concerned to the radiotherapy treatment planning. This project takes a hybrid data-oriented approach that will apply inference mechanism and expression evaluation engine to verify and accredit the External Beam Radiotherapy Treatment (EBRT) plan quality, considering the 3DCRT technique.

The aim of this project if to create a decision support system that incorporates the wealth of experience possessed by experts, policies, standards and guidelines, by applying inverse planning strategy in a hybrid data-oriented rule engine to evaluate a treatment plan (in any stage).

The main objectives sought are:

- To improve intra- and inter-planner consistency;
- To insert both the context and particular culture of each radiotherapy/oncology center in the definition of the treatment plan;
- To identify plan quality prognostic features that can be used to improve the treatment plan quality;
- To enable the development of both qualitative and quantitative metrics for plan quality evaluation and assessment

Figure 1 depicts the roles and responsibilities of ELDORADO, CEB-UNICAMP and the Clinics Hospital of UNICAMP, where the latter represents the Oncology Healthcare System for this technology transfer program. As one can observe, the ELDORADO got in charge of the project design, value proposition and software development. In turn, the CEB-UNICAMP received the task of providing both the clinical purpose and the scientific evaluation. Finally, the Clinics Hospital of UNICAMP contributed by offering the physician's intention and use case as well as the clinical evaluation.

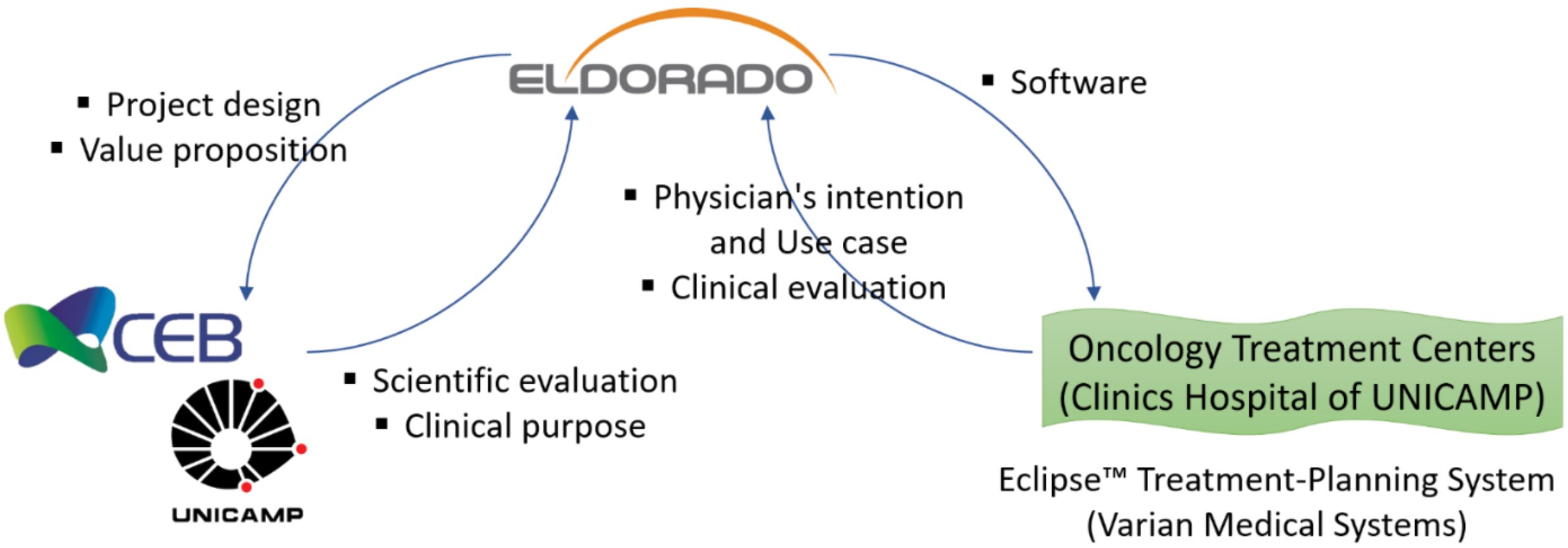

*Figure 1 - The roles and responsibilities of the Eldorado Research Institute, CEB-UNICAMP and the Clinics Hospital of UNICAMP for the project development of the proposed rule-based system during the Technology Transfer Program.*

## D. The Very Basics of Radiotherapy Treatment

Radiation therapy relies on the concept that fast proliferating cells (i.e. cancer cells) are more sensitive to ionizing radiation than healthy cells. Long story short, ionizing radiation is a form of energy derived from electromagnetic waves that carry enough energy to knock electrons from atoms or molecules. Usually, high energy x-rays, or sometimes, electrons, and more rarely protons, are used as sources for oncology treatment [8].

However, excessive radiation adversely affects all cells, including healthy tissue and critical organs. The aim of radiation therapy is to deliver a tumouricidal dose over the tumour region while minimizing the radiation received by healthy tissue and critical organs in the vicinity of a tumour or in the radiation beam path. Its application can cure many cancers by destroying the tumour or stopping it from growing any further. However, the determination of a suitable radiotherapy dose to be delivered during the treatment is a very important and complex process, since it concerns a trade-off between the expected benefit, in terms of the cancer control, and the harmful side effects, in terms of the patient survival and quality of life [8, 9].

The process from diagnosis to treatment contains many steps as one can see in Figure 2, which presents a basic radiotherapy treatment workflow. In a few words, depending on the outcomes of a medical appointment, the physician oncologist can indicates a radiotherapy treatment to the patient, and as such, the next step will be make the patient go through for an imaging scan. The most common medical imaging techniques used to the end of localizing tumours are computed tomography (CT), magnetic resonance imaging (MRI) and positron emission tomography (PET) [10]. Nonetheless, these imaging systems allow 3D reconstruction of tumours and other organs for treatment planning. In turn, a process known as image registration is needed to match the spatial information obtained by different imaging techniques.

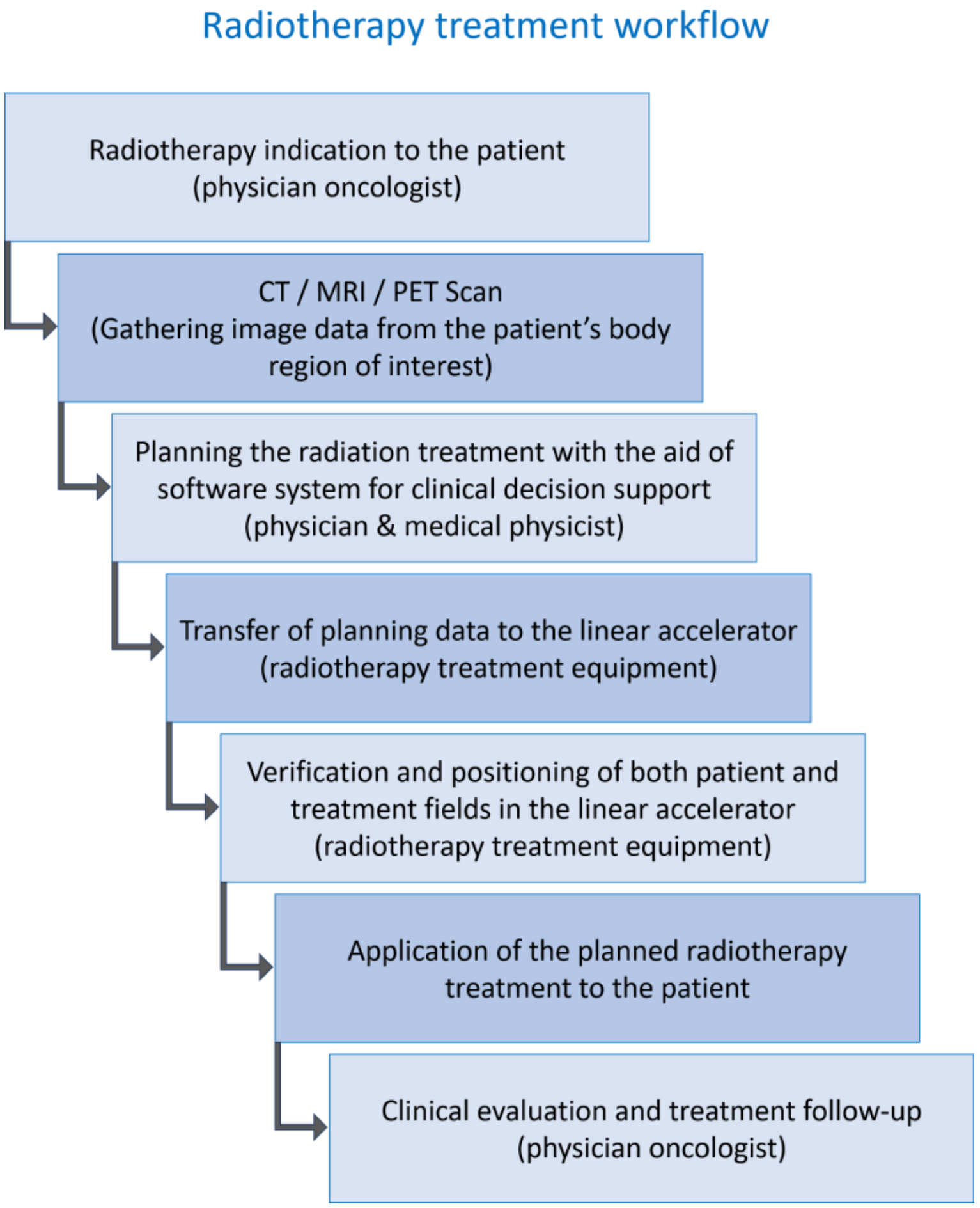

*Figure 2 - Basic diagram of a radiotherapy treatment workflow.*

In treatment planning, tumours and organs at risk are determined, and treatment parameters, such as the number and angles of beams, the modality and energy of radiation, and the intensity for each beam are chosen. Dose calculation involves models of the physical behaviour of radiation as it interacts with matter to determine dose distribution to the patient. Finally, the patient is treated, with the planned treatment delivered in fractions over a period of time, followed by clinical evaluation and treatment follow-up stages (Figure 2).

## *APPROACHES AND CHALLENGES*

Advances and improvements in medical technologies and computer hard- and software made possible the radiotherapy treatment to progress for 3D planning and, more than that allowed a computer-controlled delivery mechanism. Forward and inverse planning are two ways of approaching radiotherapy planning. The former treatment plan determines the dose distribution through interactive adjustments of the planning parameters to achieve an acceptable value, whilst the latter requires the prior specification of the dose distribution to create the treatment plan that fits the prescribed dose distribution [11].

Figure 3 shows a 3DCRT forward planning where one can see the all the common steps involved from 3D planning to its conclusion. Moreover, it makes clear how demanded is the efforts needed by the oncologist team to meet a feasible and safe radiation therapy for a given patient.

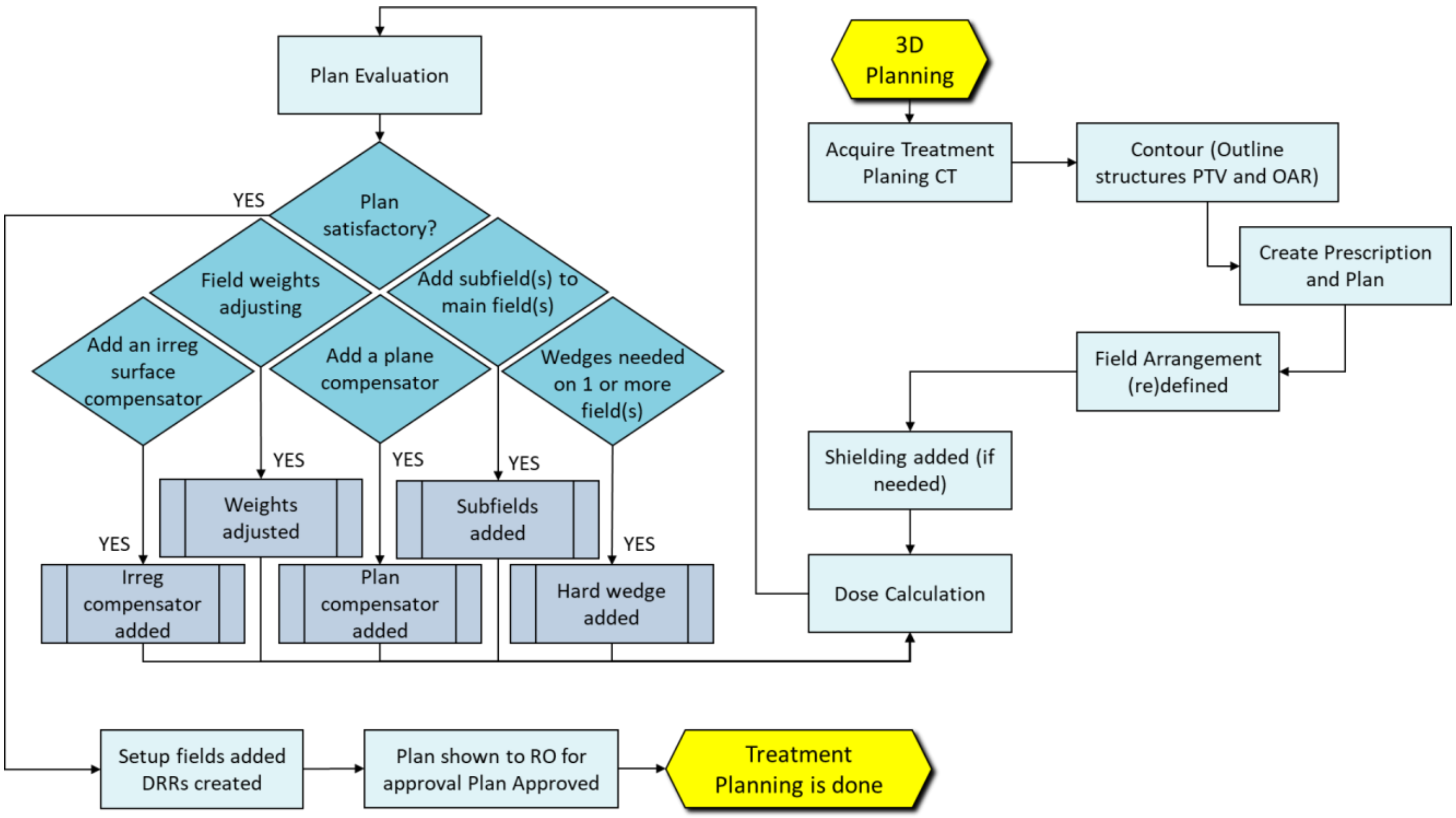

*Figure 3 – A conventional 3DCRT forward planning [Modified from the Radiotherapy Public Call Notice nº 001/2016 issued by the Ministry of Health of Brazil].*

The general process sketched by a 3DCRT forward planning follows as:

1. The oncologist views the patient's CT images and outlines the GTV (gross tumour volume), CTV (clinical target volume), PTV (Planning target volume) and OAR (organ at risk) on the images.
2. Then, the beams are placed so that they intersect at the isocenter (usually, the center of the tumour). The other parameters, such as beam weights, wedges and multi-leaf settings, are determined.
3. The dose distribution is calculated by the treatment planning system and evaluated with respect to the planning goals.
4. The beam configuration and the other parameters are modified.
5. Steps 2 to 4 are repeated until the dose distribution is satisfactory.

Normally, medical physicist shares three or four treatment plans with physician oncologists containing advantages and disadvantages of each plan. Those plans comes with the dose distribution objectives, based on the available patient information as well as their collection of image files. As a response, from the physician's feedback, the physicists may modify the treatment plan. Consequently, the outlined procedure are both time consuming and inefficient processes [12].

Depending on the cancer types, many institutions prefer to fix the dose limits to a tumour and OARs. Then, the treatment planning parameters has the goal of determining the delivered dose distribution as close as possible to the prescribed dose objectives. As a result, the inverse planning can produce radiation treatment plans that are superior to the forward planning. Notwithstanding, the generation of a good treatment plan can varies from a few hours to a few days, in complicated cases, and as such, it requires the expertise of experienced medical staff. Further, ideal dose distribution, in which no OAR and healthy tissue overdosed, and the tumour tissue is not underdosed, is impractical [13]. Additionally, as both the tumour and healthy cells interleaves themselves in the tissue, the dose allowed to be applied is limited, such that, the radiation should be appropriate to destroy the tumour cells without killing the healthy ones [14]. Besides that, the treatment plan has to (must) take into account standards, procedures, hospital policies, capabilities, recommendations and guidelines. Treatment goals also depend on other factors:

**Age of patient/Stage of cancer:** Younger patients often receives a more aggressive treatment, whereas in older or terminally ill patients, more importance to the quality of the remaining life rather than completely removing the tumour is given.

**Organs-at-risk (OAR):** Different OARs respond differently to radiation [14]. Some organs like kidney, liver, lung, among others, are known as rope organs due to their somehow sensitiveness of losing some part of itself, and as such, in a radiotherapy treatment plan, such rope organs as a whole have to be protected from uniform dosage delivered. Other organs, like spinal cord, bowel, for instance, called chain organs are also very sensitive of losing some parts of themselves, whereas they are somehow resistant to relatively high levels of radiation applied to the whole organ. As a result, during a treatment planning procedure, one needs to watch out for unusually high dosages delivered even to the small parts of such, category of organs.

Apart from that, the decision variables or the treatment plan parameters primarily concern the dose and the beam configuration and includes [12]:

- **Total radiation dose:** The radiation dose depends on the type and location of the tumour, and the total radiation is fractioned on the course over a specific time duration. This procedure is applied because healthy cells recover faster than the tumour ones, while still keeping tumour control.
- **Beam configuration:** The beam configuration is determined by the specific patient anatomy, which includes both the location and the shape of the tumour as well as the OARs in the vicinity of the target volume for treatment.
- **Number of beams:** In 3DCRT, the radiation is applied using a certain number of beams, in different directions, with the aim of reducing the total radiation dose applied to the healthy tissue in the beam's path. Usually, the number of beams can vary from 2 to 9, and its determination depends on the hospital policy for the sake of ease and effectiveness of implementation.
- **Beam weights:** The beam weights denote the intensity of each beam. Given a total prescribed radiation dose that the tumour has to receive, the individual beams can be weighted differently to make up the total dose.
- **Angle of beams:** The beams are applied at an angle to ensure that they conform to the tumour volume while avoiding as much as possible the OARs. The can be applied coplanar (all lying in one plane) or non-coplanar. There are also physical constraints about beam placement, for instance, the radiation beams should not be directed at the patient from directly underneath the treatment bed (couch).

- **Wedges:** Wedges are metallic wedge-shaped blocks, which are placed in the path of the beam to attenuate the radiation.
- **Multi-leaf collimator settings:** the leaves of the collimator shape the radiation beam.

Protocols to specify relationships between both conformal dose distributions and results of imaging studies can be derived from computable clinical guidelines and rules languages. Rules can perform the specification for radiation dose in known tumour volumes, in areas of suspected microscopic spread, while it can also determines the needed tolerances for patient motion.

Further, rules can also specify to non-target structures the maximum radiation dose is allowed to be delivered. Even more, rules can be formulated to criticizing and reviewing the conformal dose distributions generated by the medical physicists (e.g. the radiation treatment planners), in addition to potentially serves as a way of assisting in the generation of new radiation treatment plans. Lastly, rules tha would recommends conformal dose distributions during the course of a treatment could also incorporate changes in both GTV and CTV over time, along with observed radiation toxicity effects [12].

## E. A RULE-BASED SYSTEM OVERVIEW

Rule-based systems, also known as knowledge-based systems or expert systems, are an application of artificial intelligence (AI) to systems in involving human-crafted rule sets. It were conceived as a simplified way of creating some description of how to solve a problem that otherwise would be extremely complicated if were applied just a conventional algorithm development for knowledge representation of a person or group with expertise in a given field. For that reason, rules are a feasible way of encoding a human expert’s knowledge in a narrow area (or niche) into automated system. Additionally, as a sub-class of AI, knowledge representation is concerned with how knowledge is represented and manipulated, being very useful systems for reasoning, with the end of process data, to infer conclusions [15].

From that perspective, rule-based systems are software systems based on rule engine components. The term “rule engine” might be quite vague because it can be any system that uses rules, in any form, that can be applied to data to produce outcomes. Put differently, a rule engine works by evaluating collections of facts and using the results to determine new facts [16].

Essentially, rule engines are used to make inferences. Based on a validation-expression evaluation, a rule engine can generates (infer) an answer to a user’s question in accordance with its both possessed rules definitions and data that it can have access. Alternatively, a rule engine can be oriented to infer some information from a given set of data in the absence of receiving a specific question, just based on the rules contained in the engine [17].

Further, rule-based expert systems evolved from a more general class of computational models knows as production systems [18]. A production system is Turing complete, and is conceived to express propositional and first order logic in a concise, non-ambiguous and declarative manner [19]. Instead of viewing computation as a pre-specified sequence of operations, production systems view computation as the process of applying transformation rules in a sequence determined by the data. A classical production system has three major components: (1) *a global database* (working memory) that contains facts or assertions about the particular problem being solved, (2) *a rule base* (production memory) that contains the general and specific knowledge about the problem domain, and (3) *a rule interpreter* (inference engine) that carries out the problem-solving process.

The facts in the global database can be represented in any convenient formalism, such as arrays, strings of symbol, list structures, etc. The rule is a two-part structure using first-order logic for reasoning over knowledge representation, and preferably, the given statement is advantageous:

WHEN <condition> THEN <action>

Thus, consider for instance the following set of prescription facts (or findings) related to the risk of prostate cancer:

**Prescription Facts (Findings)**

- **Tumour Location:** C61.9, the code that refers to a Prostate Cancer
- **Tumour Stage:** T2aN0M0, the code sequence that give details of the lesion extension as follows:
    - *T2a:* The tumour involves one-half or one side of the prostate
    - *N0:* No regional lymph nodes
    - *M0:* No metastasis
- **PSA (Prostate-Specific Antigen):** 8ng/ml
- **Gleason Score:** 6, which is the grading system used to determine the aggressiveness of prostate cancer – typical range between 6-10)

Based on the gathered set of facts above, one can establish a rule with a [WHEN] and [THEN] structure as follows:

- Rule 1: **[When]** the Tomour Location is C61.9 (Prostate), and the Primary Tomour Stage T is T1a, T1b, T1c or T2a, and the Lymph Node Stage is N0, and the Metastasis Stage is M0, and PSA $<$ 10ng/ml, and Gleason Score $\leq$ 6, **[Then]** Classify the Tumour as a Low Recurrence Risk Prostate Cancer.
- Rule 2: **[When]** Low Recurrence Risk Prostate Cancer, **[Then]** apply 3D-CRT or IMRT with dose of 75 to 79.2Gy up to 44 fractions with maximum dose per fraction of 1.8Gy.

Hence, Figure 4 depicts a basic diagram example that illustrates parts of a rule-based system using a set of data under certain conditions to obey some rules.

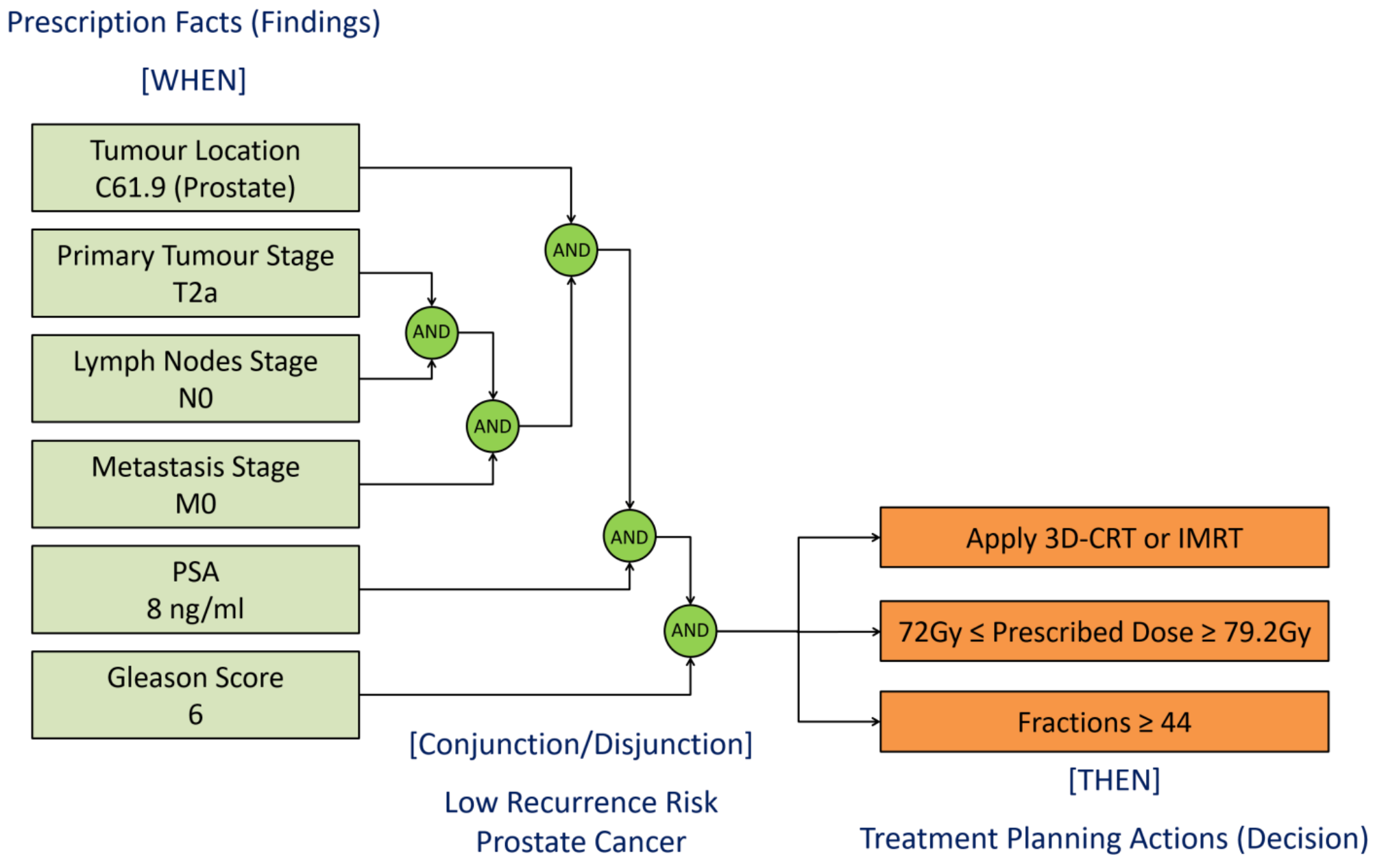

*Figure 4 – Diagram showing the basics of a rule-based system.*

In general, the condition part or sensory precondition [WHEN], the left-hand-side (LHS) of a rule can be any pattern that can be matched against the database. Once a match is achieved, the action part [THEN], the right-hand-side (RHS) of the rule can be executed. Such action can be also any arbitrary procedure employing the bound variables.

In turn, the rule interpreter, or inference engine, has the task of deciding which rules to apply. The rule interpreter for a classical production system executes in a “recognize-act” cycle. It turns out that the rule interpreter cycles through the condition parts of the rules, looking for one that matches the current database, and executes the associated actions for (some or all) rules that satisfy that condition.

Typically, the inference engine, as shown in Figure 5, has three components: **Pattern Matcher**, **Agenda** and **Execution Engine**. The **Pattern Matcher** compares the data of rules and facts and adds the rules that satisfy the facts into Agenda. In turn, **Agenda** manages the execution sequence of the rules which Pattern Matcher chooses, and **Execution Engine** executes these rules. A system with a large number of rules and facts may result in many rules being true for the same fact assertion; these rules are said to be in conflict. The Agenda manages the execution order of these conflicting rules using a Conflict Resolution strategy.

Forward chaining and backward chaining are primarily the two modes of operation of inference engines. The former starts with the available data and uses inference rules to extract more data until reaching a goal. Thus, an inference engine that uses a forward chaining, search the inference rules until it finds one where the antecedent (LHS) is known to be true. When such rules are found, the engine can make a conclusion or infer the consequent (RHS), which results in the addition of new information to its data. This iterative process will happen until to reach an established goal [19].

On its part, backward chaining starts with a list of goals (or hypothesis) and works, as its name indicates, backwards from the consequent to the antecedent to verify whether any data supports any of these consequents or not. An inference engine that uses a backward chaining would search the inference rules until it finds one with a consequent (RHS) that matches an established goal. If the antecedent (LHS) of that rule is unknown to be true, then it is added to the objective list (i.e. for one’s goals to be confirmed one must also provide data that confirms this new rule) [20]. Table 1 summarizes the pros and cons between forward and backward chaining [21].

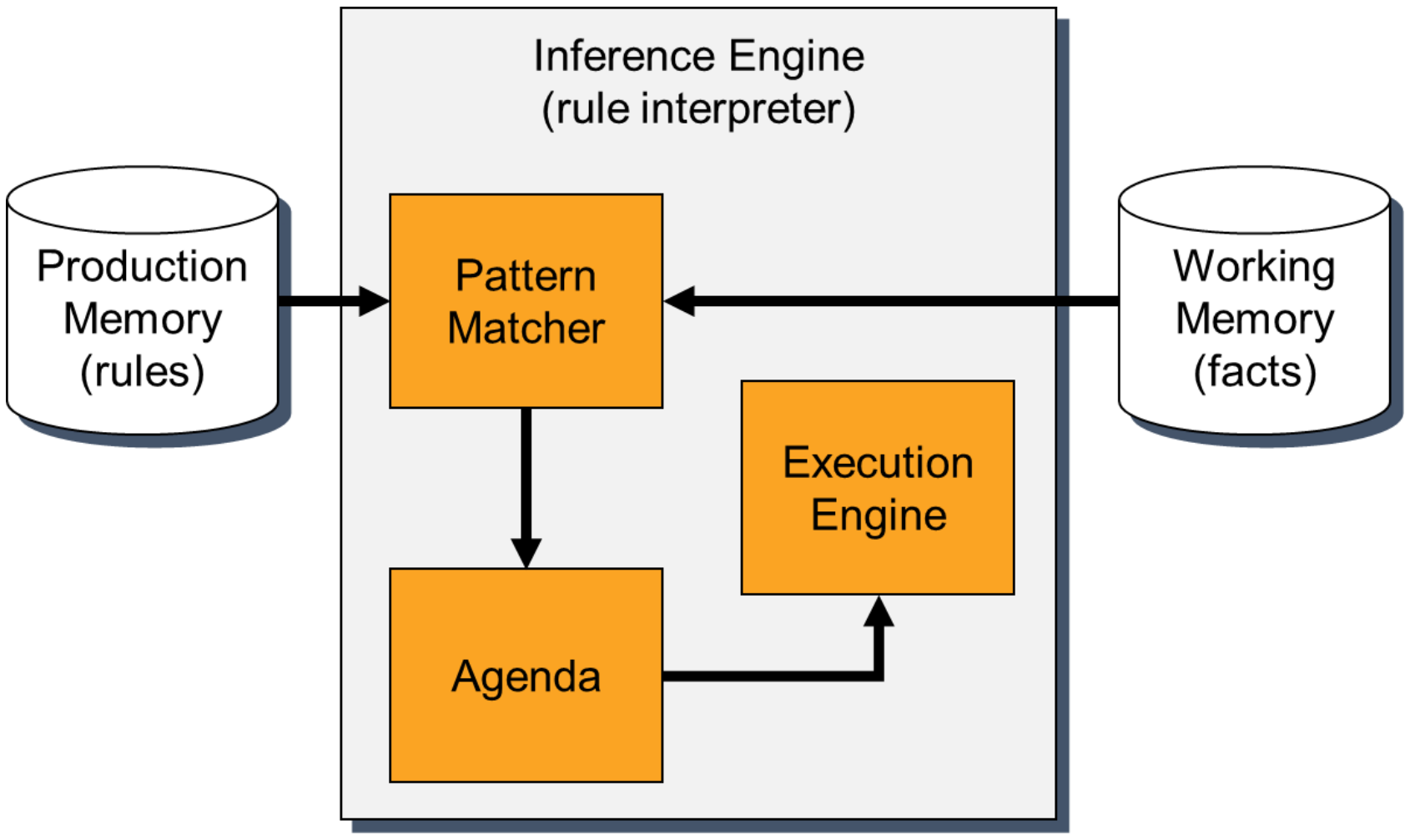

*Figure 5 - The components of an inference engine (rule interpreter) [Modified from [34]].*

***Table 1 - Comparison between forward and backward chaining [21]***

| Forward chaining | Backward chaining |
|---|---|
| Starts with the initial facts | Starts with some hypothesis or goal |
| Asks many questions | Asks few questions |
| Tests all the rules | Tests some rules |
| Slow | Fast |
| Provides a huge amount of information from just a small amount of data | Provides a small amount of information from just a small amount of data |
| Attempts to infer everything possible from the available information | Searches only that part of the knowledge base that is relevant to the current problem |
| Primarily data-driven | Goal-driven |
| Uses inputs; searches rules for answer | Begins with a hypothesis; seeks information until the hypothesis is accepted or rejected |
| Top-down reasoning | Bottom-up reasoning |
| Works forward to find conclusions from facts | Works backward to find facts that support the hypothesis |
| Tends to be breadth-first | Tends to be depth-first |
| Suitable for problems that start from data collections | Suitable for problems that starts from a hypothesis |
| Non-focused; it infers all conclusions from data, may answer unrelated questions | Focused; questions all focused to prove the goal and search as only the part of knowledge-base that is related to the problem |
| Explanation not facilitated | Explanation facilitated |
| All data is available | Data must be acquired interactively, i.e., on demand |
| A small number of initial states but a high number of conclusions | A small number of initial goals and a large number of rules match the facts |
| Forming a goal is difficult | Easy to form a goal |

All the same, while both forward and backward chaining rules engines can, in some cases, provide powerful reasoning capabilities for arriving at solutions, the time taken to get such answers is difficult to predict. These rules engines belong to the category of non-deterministic ones. Differently, the third class of rule engine known as deterministic utilize domain-specific language (DSL) approaches for better describing policies. Thus, deterministic rules engines have more predictable and consistent behavior, and as such, are suited for applications that intend to implement a stated set of policy rules [20].

## F. Knowledge Representation Model

Rule-based systems, an applied AI in a knowledge specific domain, makes the knowledge representation model the most important ingredient for developing an AI system. Hence, knowledge representation and reasoning concerns to representing information about a specific domain such that a computer system can utilize to solve complex tasks, such as diagnosing a medical condition or the planning of a 3D radiotherapy treatment for targeting a tumour.

It turns out that the inference engine needs a computationally compliant format to represent the incoming data per a set of rules [22, 23, 24]. The format of such rules as used by a specific rule engine can be set out as the knowledge representation model. Examples of knowledge representation models commonly employed in the biomedical and computer science domains, just to mention a few, includes: context-specific Boolean rules, suchlike those set in the Mycin clinical decision-support system [25, 26], complex axiomatic statements like those employed by the Arden syntax for medical logic modules [27, 28, 29], and frames such as those used in numerous biomedical ontologies and intelligent agents [23, 24, 30, 31].

In defiance of many advances in both knowledge representation and inference technologies approaches, the production rules remain the most prevailing because it allows express first-order logic declaratively in an unambiguous human readable form, whilst keeping machine interpretability. Table 2 depicts an overview of the advantages and disadvantages of applying production rules mechanisms. Consequently, despite its procedural role in systems, essentially, rules are declarative representations [32].

***Table 2 - Advantages and disadvantages of declarative representation [32]***

| Advantages of production rules | Disadvantages of production rules |
|---|---|
| **Compact representation of general knowledge**: Rules can easily represent general knowledge about a problem domain in autonomous, relatively small chunks.<br><br>**Naturalness of representation**: Rules are a very natural knowledge representation method, with a high level of comprehensibility.<br><br>**Modularity**: Each rule is a discrete knowledge unit that can be inserted into or removed from the knowledge base, without taking care of any other technical detail (as long as other rules are not affected).<br><br>**Provision of explanations**: The ability to provide explanations for the derived conclusions in a straightforward manner. | **Knowledge acquisition bottleneck**: The standard way of acquiring rules through interviews with experts is cumbersome and time-consuming.<br><br>**Brittleness of rules**: It is not possible to draw conclusions from rules when there are missing values in the input data.<br><br>**Inference efficiency problems**: In certain cases, the performance of the inference engine is not the desired one, especially in very large rule bases.<br><br>**Difficulty in maintenance of large rule bases**: The maintenance of rule bases is getting a difficult process as the size of the rule base increases. The rule base may have problems such as, redundant rules, conflicting rules, rules with redundant or missing conditions, missing rules, etc.<br><br>**Problem solving experience is not exploited**: A rule-based system is not self-updatable, in the sense that there is no inherent mechanism to incorporate experience acquired from dealing with past problems.<br><br>**Interpretation problems**: The general nature of rules may create problems in the interpretation of their scope during reasoning. |

With this in mind, one of the most comprehensive knowledge representations in clinical terminology in use around the world is the one called SNOMED CT, which stands for Systematized Nomenclature of Medicine -- Clinical Terms, and was the chosen one adopted in our project. Among its benefits, one can noticed the following, extracted from their own website [33]:

a. Electronic health records that allows the access to relevant and critical clinical information, in addition to, the increasing opportunities for real time decision support to more accurate retrospective reporting for research, data analytics, precision medicine and management;
b. Benefits for individuals, as it removes language barriers offering multi-language support, it also allows accurate and comprehensive analysis that identify patients who require follow-up or changes of treatment improve communication, among others;
c. Benefits for population, by allowing early identification of emerging health issues, monitoring of population health and agile response to changing clinical practices. Additionally, it enables accurate access to relevant information, reducing costly duplications and errors, just to mention a few;
d. Evidence-based healthcare, because it enables links between clinical records and clinical guidelines, can enhances the quality of care, raising its cost-effectiveness. Further, it can limit the frequency and impact of adverse healthcare, just to citing some of the benefits.

As an illustration of the power of SNOMED CT system for knowledge representation in cancer data, Figure 6 depicts the net of terminologies associated with carcinoma of breast (based on the International Classification of Diseases number 10, ICD-10). As one can observe, the presented ontology net connects the disorders with the associated findings, allowing the establishment of a huge and complex interconnection of knowledge for that specific type of cancer.

### *KNOWLEDGE REPRESENTATION ISSUES*

For specific problems, regardless of the knowledge representation language chosen, may be essential to represent it and reason on it with spatial models, temporal relations, compound objects, possible worlds, beliefs and expectations. Hence, issues suchlike consistency, completeness, robustness and transparency are equally important in the construction of a knowledge base for expert systems, and as such, plays the major design considerations functions [18]. Therewithal, still according to the authors of [18]:

Much of the knowledge flowing out of an expert is uncodified and comes with uncertainty. Then, it is unrealistic to assume that the knowledge base can be sufficiently cleansed to withstand a logician's scrutiny. As a result, consistency in the knowledge base is obviously desirable.

On its part, syntactic and semantic completeness are both important issues that must be taken account of in knowledge representation. The former refers to a logical requirement that many rule-based languages fail to satisfy (e. g. assertions as quantified statements that are difficult or impossible to express), whereas, the latter, refers to the meaning of symbols, and as such, it will almost certainly fail to cover some interesting (sometimes important) possibilities because the cost of checking all combinations for completeness is prohibitive. Thus, the best way of checking the completeness of rules coverage made by an inference system is by choosing carefully the test cases.

In turn, capability of representing properly the degrees of imprecision is an important part of every representation methodology, especially because there would be a temptation of making overly precise assertions for the knowledge base, even when there is no justification for fine precision. Consequently, precision in specialized domains is feasible for many of the facts and rules, but certainly it will not achieve for all. At the same time, default knowledge is an important protection against incompleteness, but generally, it requires stating the default for each class of actions explicitly.

Typically, causal models provide a detailed specification of how complex devices works, whether it be biological or mechanical. For its part, temporal relations, as causal ones, are still generally difficult to represent and use in satisfactory ways. Lastly, although strategies for problem-solving are a very important part of expertise, they are also difficult to represent and to use it efficiently.

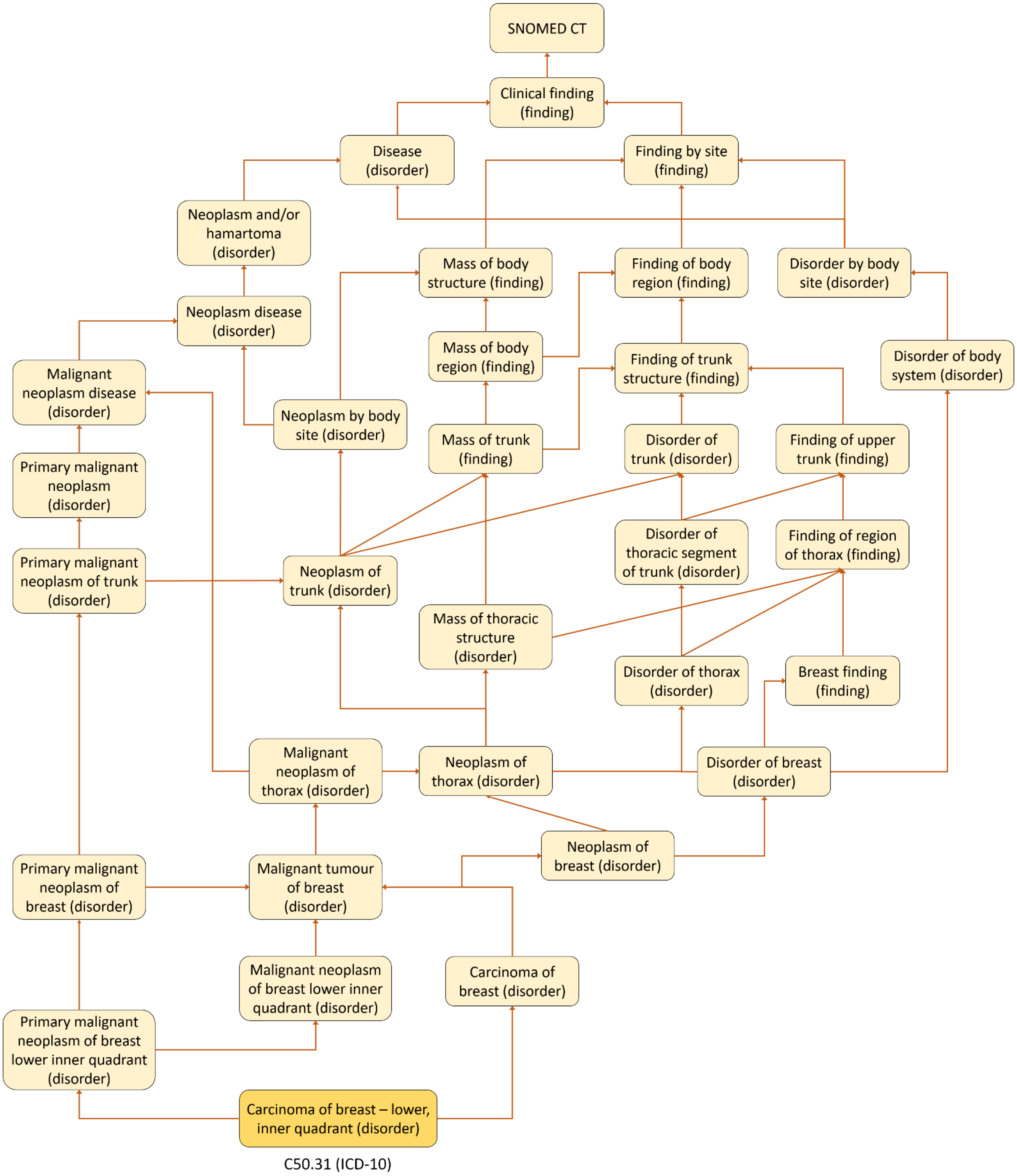

*Figure 6 - Example of a knowledge representation model of a breast carcinoma based on the SNOMED CT system [Modified from the "NCI Workshop: The Role of Ontology in Big Cancer Data Session 3: Cancer big data and the Ontology of Disease Bethesda, Maryland May 13, 2015" – PDF document].*

### *DECISION TABLE*

Decision tables are a way of structuring conditional logic into a tabular form. In this context, conditional logic relates to a set of tests that will give rise of a set of actions. Four quadrants are one of the most common shapes of a decision table used to organize conditional logic as one can see in Figure 7. The upper two quadrants named the condition stub and condition entry, describe conditions for which logic must be tested. Thus, condition stub represents the list of conditions or tests, whilst condition entry indicates for each column what results from each condition is necessary for this column to execute.

Conversely, the lower two quadrants describe the actions to be taken depending on the outcome of the condition tests together with the associated actions. The action stub label represents the list of actions to be performed and the action entry field provides, for each column, an “X” mark for each action that should be executed.

| | Rule 1 | Rule 2 | . . . . . |
|---|---|---|---|
| Condition Stub | Condition Entries | | |
| Action Stub | Action Entries | | |

*Figure 7 - Structure of a decision table.*

In the end, a decision table is considered balanced or complete if it includes every possible combination of input variables (i.e. balanced decision tables prescribe an action in every situation where the input variables are provided. With balanced decision tables, the user defines “every” path through the logic, while with unbalanced ones, the user defines only a subset of paths through the logic. Thence, unbalanced decision tables reduce the clutter and complexity even further, and as such, enhance the table ability to represent the logic behind the table. Additionally, unbalanced tables, in favor of documenting the policy more clearly, sacrifice, in response, the goal of documenting every path through the logic [34].

## G. THE PROPOSED RULE-BASED SYSTEM ARCHITECTURE

The proposed rule-based architecture comprises two layers, named as rule engineering layer and a fact extraction layer. Figure 8 depicts the diagram block and the components of the system. The rule engineering layer has five main features: Rule editor, Ontology manipulator, MLM validator, MLM2Code translator, MLM repository, inference mechanism.

- The rule editor provides an user interface for physicians to creating and editing sharable rules and treatment planning evaluation criteria in a Ripple-Down Rules (RDR) fashion, using Arden Syntax MLM constructor and decision tables, based on concepts and terminology from a Domain Specific Ontology – including/acquired from NCCN Guidelines, RTOG, QUANTEC, ICD-O and TNM standards, among others – and from SNOMED CT.
- The ontology manipulator extracts the domain concepts and maps it to an instruction set.
- MLM validator verifies the syntax of rules according to the Arden Syntax, and both rule consistency and compatibility (e.g. duplication, conflict).
- MLM2Code translator transform the standard MLM into executable classes.

- The MLM repository stores valid MLMs and classes (mapping).
- Differently from [35], the proposed inference mechanism has four components: **pattern matcher**, **agenda**, **execution engine**, and an **evaluation criteria invoker**. The pattern matcher compares the patient facts and prescription rule data and adds the rules which satisfy the facts into agenda. Agenda manages the execution sequence of the rules which pattern matcher chooses, and execution engine execute these rules. The execution engine loads guidelines, similar cases, and the decision tables/associated rules into the user interface. The user interacts with the evaluation criteria invoker, using the decision tables/associated rules loaded (questions) that requests evidence to the data extraction layer and returns an answer (fail/pass).

In turn, the fact extraction layer is responsible for acquiring facts, data and attributes from a database. This layer has two components, namely fact provider and data manipulator.

- Fact provider is a middleware that queries and manages data acquisition, processing and transformation services based on the fact requested by the evaluation criteria invoker component.
- The data manipulator processes and manipulates data from database using specialized application and tools, when requested by the fact provider.

Hence, the key architectural features of the proposed system can be summarized as follows:

- **Passive:** The engine only executes decisions when explicitly invoked by the user.
- **Stateless:** The engine does not store state or data between activations.
- **Executable:** All decisions are compiled into an executable code.
- **Deterministic:** For a given input and rule set, the path to derive the output is defined.
- **Standardize Knowledge Base:** Knowledge base realized by using standard knowledge representation units of Medical Logic Modules (MLM), using HL7 Arden Syntax, and SNOMED CT and ICD-O standard terminology.

### *Evaluation Criteria Classification*

To allow partial verification of a treatment plan, under a specific aspect, the rules for evaluation and accreditation should be segmented into classes, as follows:

- **Preconditions:** Check the adherence of the preconditions to the specific ontology and the completeness of the set of preconditions and data associated with the definition of the applicable set of rules.
- **Conventions and nomenclatures:** Include rules associated with the specific domain dictionary and the conventions for readability and uniformity of the plan.
- **Regions, structures and organs at risks:** Rules and recommendations (guides) on the regions and structures that should be outlined in the particular treatment. Among the rules are the obligation to delineate a particular organ or the existence of a region; the limits of a structure date, for example the applicable minimum volume, or its relationship with other structures, etc. For the recommendations (which cover the rules that require active intelligence for its verification), are the anatomical location of a region date, or its extension.
- **Constraints and dosage:** Rules associated with the adherence of the plan to the prescription and distribution of radiation by body, structures and regions. QUANTEC recommendations, T5/5 and T50/5 are examples of this class of rules.

- **Quality criteria:** The rules in this class verify a finalized plan regarding pre-established criteria for its quality. Number and extent of hot spots and cold spots, homogeneity and compliance index, dose obtained from monitor units from machine calibration data, etc.

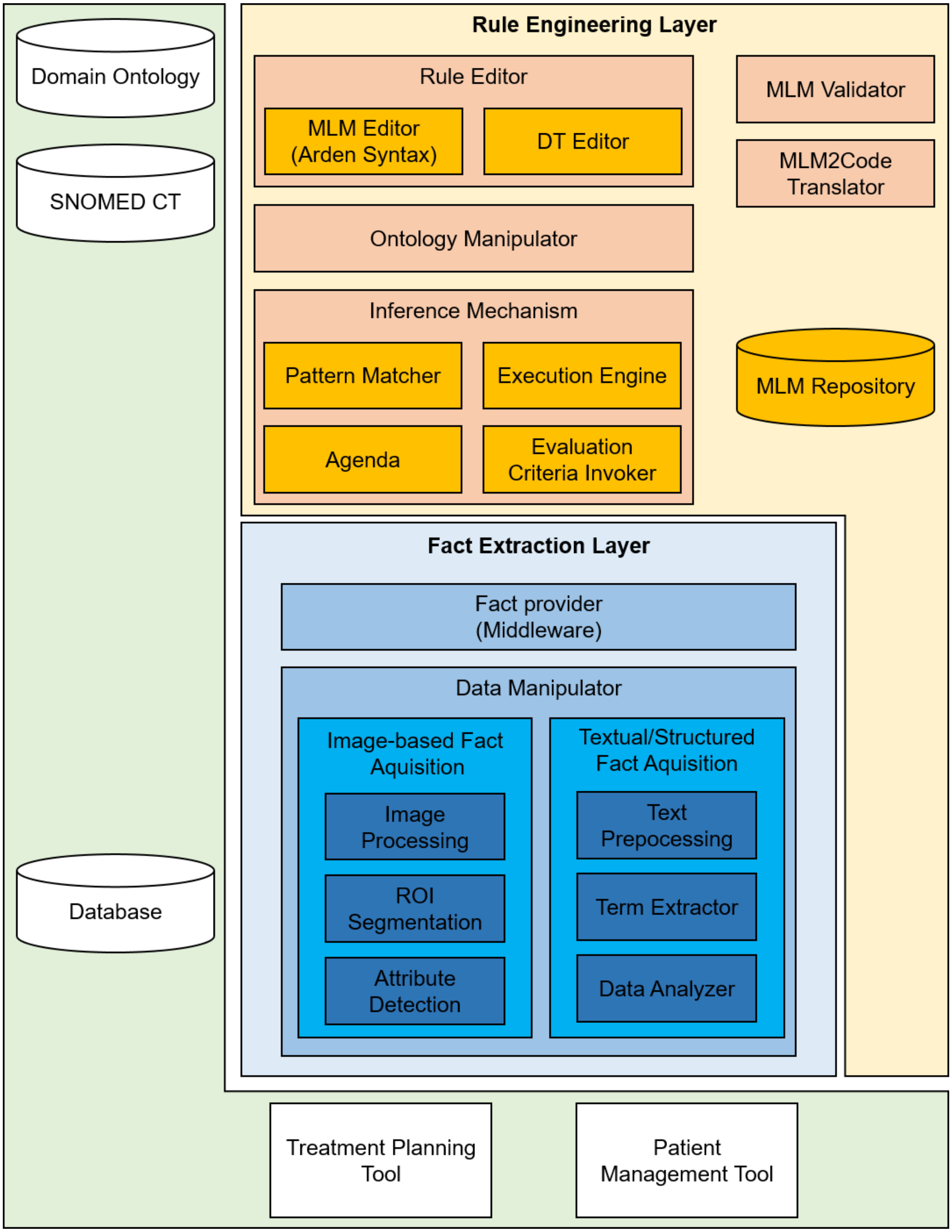

*Figure 8 - The proposed architecture for the rule-based system to be used in radiotherapy treatment planning.*

### *Patient Data*

The data from the patient becomes eligible to be stored in the system only when a primary tumour diagnosis has been confirmed, which means that the database should only reflect the completed medical diagnosis. The patient data is described by specific characteristics of a tumour, associated radiation therapy treatments and examinations, which can include temporal and factual data.

In turn, tumour-specific characteristics lie in either primary tumour diagnosis, local and opposite mammary recurrences (as the ICD-0 and TNM codes) or metastatic recurrence (as location). Treatments includes a prescribed dosage of radiation, its fractions, radiation beam path, and so on. In addition, examinations are represented by a collection of medical images, Gleason scores, blood and urine tests, just to mention a few [12].

Ultimately, it is expected that the patient data should be stored in structured fields, using the domain-specific ontology, in order to speed up the patch matcher process.

### *Workflow*

The workflow represents how the proposed architecture should behave. Thus, it is provided an overview of how the data ought to be treated by this system.

First, the system determines the applicable rules for a particular prescription. Thus taking up the example of prostate cancer prescription as shown in Figure 4, rewritten here for convenience, we have:

**Prescription Facts (Findings)**

- **Tumour Location:** C61.9, the code that refers to a Prostate Cancer
- **Tumour Stage:** T2aN0M0, the code sequence that give details of the lesion extension as follows:
  - *T2a:* The tumour involves one-half or one side of the prostate
  - *N0:* No regional lymph nodes
  - *M0:* No metastasis
- **PSA (Prostate-Specific Antigen):** 8ng/ml
- **Gleason Score:** 6, which is the grading system used to determine the aggressiveness of prostate cancer – typical range between 6-10)

In this step, the system acts as an expert system and tries to find the set of rules applicable to the facts. In that case:

- <u>Rule:</u> **[When]** the Tomour Location is C61.9 (Prostate), and the Primary Tomour Stage T is T1a, T1b, T1c or T2a, and the Lymph Node Stage is N0, and the Metastasis Stage is M0, and PSA < 10ng/ml, and Gleason Score ≤ 6, **[Then]** CLASSIFY the Tumour as a Low Recurrence Risk Prostate Cancer and LOAD Low Recurrence Risk Prostate Cancer EVALUATION CRITERIA, GUIDELINES, CASES, etc.

The system then operates as a validation/expression evaluation engine. In other words, in a pass/fail fashion) A user interacting with the Evaluation Criteria Invoker can ask specific questions stated by the set of evaluation criteria organized into classes (Preconditions; Conventions and nomenclatures; Regions, structures and OARs; Constraints and dosage; Quality criteria). Some examples are:

- **Precondition rule (Low Recurrence Risk Prostate Cancer):**
  - The treatment technique should be 3DCRT or IMRT
  - The total dose should be ≥ 75 and ≤ 79.2 Gy
  - The number of fractions should be ≥ 44
  - The maximum dose per fractions should be ≤ 1.8Gy

- **Conventions and nomenclatures:**
  - The contour of the CTV (Clinical Target Volume) should be dark blue
  - The contour of the GTV (Gross Target Volume) should be red
  - The structure name should obey the domain ontology
- **Regions, structures and OARs rule (Low Recurrence Risk Prostate Cancer):**
  - The rectum should be delineated
  - The bladder should be delineated
  - 45% of the small bowel volume should be < 195cc (entire potential space within peritoneal cavity)
- **Constraints and dose rule (Low Recurrence Risk Prostate Cancer):**
  - 80% of the bladder volume should be irradiate with less than 15% of the prescription dose (V80 < 15%) OR 65% of its volume should be irradiate with less than 17% of the prescription dose (V65 < 17%)
  - 100% of the PTV should be irradiate with more than 98% of the prescription dose (V100 > 98%)
  - The maximum point dose in the PTV (Planning Target Volume) should be less than 107% of the prescription dose
  - The mean dose to the penile bulb should be < 52.5Gy.
- **Quality criteria rule (Low Recurrence Risk Prostate Cancer):**
  - The conformity index should be < 1.4
  - The number of cold spots (regions within PTV with dose ≤ 95% of the prescription dose) should be zero
  - The homogeneity index should be > 0.98

In some cases, the question cannot be answered directly by the data in the treatment plan, as for instance in the determination of the conformity index and the homogeneity index. In that case, the rule engine request specialized/advanced data processing to fulfill the facts needed.

In other cases, active intelligence is required to answer the question. As an illustration, the anatomical determination of the nodal CTV boundaries (prostate cancer), which should have superior margin greater than the junction of outer and inner iliac vessels, but not superior to the junction of common iliac vessels. In this case, it cannot be verified without the use of an active intelligence (natural or artificial) to distinguish and determine the compliance with this particular rule. For rules that require active intelligence, the system provides the rule and guideline, and in response, requests the user to confirm the plan's compliance.

## CONCLUSION AND FUTURE WORK

It was presented the concepts of a proposed rule-based system aimed to be designed and implemented as a software system solution to aid Oncology Healthcare Institutions in Brazil suchlike the Clinics Hospital of UNICAMP. The main conception behind of its proposal is to offer somehow a way of simplify the user workflow of 3D radiotherapy treatment planning, and additionally, provide a way of improve the quality assessment of the plans for the oncologists team, helping them in evaluating and decision-making process, reducing in a certain sense, the planning mistakes possibly made in clinical practice.

Notwithstanding, the proposed solution is applicable in Big Data applications, and therefore, expansible to the Brazilian Public Health System as a whole, as it is possible to combine the rule engine with the machine learning field. The conceived architecture can be used within the specific context and reality of each existent Oncology Treatment Center in Brazil as a whole. More than that, the desired and intended application of this software system are to fill, in the future, the gap between the renowned

Brazilian centers of reference for cancer treatment and those smaller and more distant centers located in the countryside of Brazil, where it is known that there is a shortage of more qualified professionals.

As a result, the idea will be to provide to those limited centers access to clinical practices practiced by those reference centers in order to create a national standard radiotherapy treatment planning for entire Brazil.

# References

[1] Ma, Xiaomei and Herbert Yu. “Global burden of cancer” *Yale journal of biology and medicine* vol. 79, 3-4 (2007): 85-94.

[2] GBD 2015 Risk Factors Collaborators. “Global, regional, and national comparative risk assessment of 79 behavioural, environmental and occupational, and metabolic risks or clusters of risks, 1990-2015: a systematic analysis for the Global Burden of Disease Study 2015”. *Lancet Glob Health*. 2016 Oct; 388 (10053):1659-1724.

[3] Plummer M, de Martel C, Vignat J, Ferlay J, Bray F, Franceschi S. “Global burden of cancers attributable to infections in 2012: a synthetic analysis”. *Lancet Glob Health*. 2016 Sep; 4(9):e609-16.

[4] Araújo LP, Sá NM, Moraes AT. “Necessidades Atuais de Radioterapia no SUS e Estimativas para o Ano de 2030”. *Revista Brasileira de Cancerologia*, 2016; 62(1): 35-42.

[5] Yi lu Yang and Qiang Liu, "3D CT simulation and treatment planning system for radiotherapy," *International Conference on Information Acquisition*, 2004. Proceedings, Hefei, 2004, pp. 436-439.

[6] Lobach, David F et al. “Increasing Complexity in Rule-Based Clinical Decision Support: The Symptom Assessment and Management Intervention” *JMIR medical informatics*. Vol. 4(4) e36. 8 Nov. 2016.

[7] Grosan C., Abraham A. “Rule-Based Expert Systems”. In: *Intelligent Systems. Intelligent Systems Reference Library*, Vol 17. (2011) Springer, Berlin, Heidelberg

[8] Bortfeld, T and R Jeraj. “The physical basis and future of radiation therapy”. *British Journal of Radiology*. Vol. 84, 1002 (2011): 485-98.

[9] Seo, Songwon et al. “Radiation-related occupational cancer and its recognition criteria in South Korea”. *Annals of occupational and environmental medicine*. Vol. 30 (9), Feb 2018.

[10] Van den Berge DL, De Ridder M, Storme GA. “Imaging in radiotherapy”. *Eur J Radiol*. Oct 2000. 36(1):41-8.

[11] Petrovic, Sanja; Khussainova, Gulmira; Jagannathan, Rupa. “Knowledge-light Adaptation Approaches in Case-based Reasoning for Radiotherapy Treatment Planning”. *Artificial Intelligence in Medicine*, Vol. 68, p. 17-28, March 2016.

[12] Jagannathan, Rupa. “A Case-based Reasoning System for Radiotherapy Treatment Planning for Brain Cancer”. PhD thesis, University of Nottingham, July 2013. https://eprints.nottingham.ac.uk/29318/1/595290.pdf

[13] Hamacher, H. W., Kuefer, K. H. “Inverse Radiation Therapy Planning – A Multiple Objective Optimization Approach”. *Discrete Applied Mathematics*, Vol. 118, p. 145-161. 2002.

[14] Holder, A. “Operations Research and Heath Care: A Handbook of Methods and Applications”. In: Brandeau, M. L.; Sainfort, F.; Pierskalla, W. P. (eds). Kluwer Academic Publishers. 2004.

[15] L. Alty and G. Guida. “The Use of Rule-based System Technology for the Design of Man-Machine Systems”. *IFAC Proceedings Volumes*, Vol. 18(10), p. 21-41, Sep 1985.

[16] https://docs.jboss.org/drools/release/5.3.0.Final/drools-expert-docs/html/ch01.html, accessed on Nov 19th of 2018.

[17] Chisholm, Malcolm. "How to Build a Business Rules Engine: Extending Applications Functionality through Metadata Engineering". San Francisco, USA: Morgan Kaufmann (Elsevier), 2004.

[18] Buchanan, Bruce G.; Duda, Richard O. "Principles of Rule-Based Expert Systems". Report No. STAN-CS-82-926, Department of Computer Science, Stanford University, Stanford, USA, 1982.

[19] Feigenbaum, Edward A.; McCorduck, Pamela; Nii, H. Penny. "The Rise of the Expert Company: How Visionary Companies are using Artificial Intelligence to Achieve Higher Productivity and Profits". London, UK: MacMillan, 1988.

[20] Russel, Stuart; Norvig, Peter. "Artificial Intelligence: A Modern Approach". Second Edition. New Jersey, USA: Prentice Hall, 2003.

[21] Al-Ajlan, Ajlan. "The Comparison between Forward and Backward Chaining". *International Journal of Machine Learning and Computing*, Vol. 5, No. 2, p. 106-113, April 2015.

[22] Slatz, Joel; Niland, Joyce; Payne, Philip; Shah, Hemant; Stahl, Douglas. "Rules Engine Technologies Across caBIGtm Workspaces". *Technical Report*, 2007. http://proteme.org/papers/pdf/caBIG_Rules_Engine_WP_v4.pdf

[23] Jose-Luis, A., et al., "Using a high-level knowledge representation for expert systems knowledge acquisition and prototyping" In: *Proceedings of the 1993 ACM/SIGAPP Symposium on Applied Computing: states of the art and practice*. 1993, ACM Press: Indianapolis, Indiana, United States.

[24] Lavrac, N. and I. Mozetic, "Methods for knowledge acquisition and refinement in second generation expert systems". *ACM SIGART Bulletin*, 1989(108): p. 63-69.

[25] Shortliffe, E.H., et al., "Computer-based consultations in clinical therapeutics: explanation and rule acquisition capabilities of the MYCIN system". *Comput Biomed Res*, 1975. 8(4): p. 303-320.

[26] Wraith, S.M., et al., "Computerized consultation system for selection of antimicrobial therapy". *Am J Hosp Pharm*, Vol, 33(12): p. 1304-8, 1976.

[27] Hripcsak, G., "Arden Syntax for Medical Logic Modules". *MD Comput*, 1991. 8(2): p.76, 78.

[28] Hripcsak, G., et al., "Rationale for the Arden Syntax". *Comput Biomed Res*, 1994. 27(4): p. 291-324.

[29] Johansson, B. and L. Bergvin, "Arden Syntax as a standard for knowledge bases in the clinical chemistry laboratory". *Clin Chim Acta*, 1993. 222(1-2): p. 123-128.

[30] Cimino, J.J., "From data to knowledge through concept-oriented terminologies: experience with the Medical Entities Dictionary". *J Am Med Inform Assoc*, 2000. 7(3): p. 288-297.

[31] Cimino, J.J., et al., "Knowledge-based approaches to the maintenance of a large controlled medical terminology". *J Am Med Inform Assoc*, 1994. 1(1): p. 35-50.

[32] Prentzas, Jim; Hatzilygeroudis, Ioannis. "Categorizing Approaches Combining Rule-Based and Case-Based Reasoning". *Expert Systems* 24(2):97-122. May 2007.

[33] https://www.snomed.org/, accessed on Nov 19th of 2018.

[34] Snow, Paul. "Decision Tables". *DTRules: A Java Based Decision Table Rules Engine.* http://www.dtrules.com/newsite/?p=90

[35] Di Liu, Tao Gu and Jiang-Ping Xue, "Rule Engine based on improvement Rete algorithm". *The 2010 International Conference on Apperceiving Computing and Intelligence Analysis Proceeding*, Chengdu, 2010, pp. 346-349.